\documentclass[useAMS,usenatbib]{mn2e}

\usepackage{graphicx}
\usepackage{times}
\usepackage{natbib}
\usepackage{epsfig}

\newcommand{\be}{\begin{equation}}
\newcommand{\ee}{\end{equation}}
\newcommand{\hp}{HEALPIx}
\newcommand{\dd}{{\rm d}}

\title[
The imprints of local superclusters on the SZ signals and their
detectability with Planck
]
{
The imprints of local superclusters on the Sunyaev-Zel'dovich 
signals and their detectability with Planck
}
\author[K. Dolag et al.]
{
K. Dolag$^{1}$\thanks{E-mail: kdolag@mpa-garching.mpg.de}, 
F.K. Hansen$^2$\thanks{E-mail: f.k.hansen@astro.uio.no}, 
M. Roncarelli$^{3}$\thanks{E-mail: mauro.roncarelli@studio.unibo.it}, 
L. Moscardini$^{3}$\thanks{E-mail: lauro.moscardini@unibo.it}\\
$^1$ Max-Planck-Institut f\"{u}r Astrophysik,
Karl-Schwarzschild-Stra\ss{}e 1, D-85740 Garching
bei M\"{u}nchen, Germany\\
$^2$ Institute of Theoretical Astrophysics, University of Oslo,
P.O. Box 1029 Blindern, N-0315 Oslo, Norway\\
$^3$ Dipartimento di Astronomia, Universit\`a di Bologna,
via Ranzani 1, I-40127 Bologna, Italy }

\begin{document}

\date{Accepted ???. Received ???; in original form May 2005}

\pagerange{\pageref{firstpage}--\pageref{lastpage}} \pubyear{2005}

\maketitle
\label{firstpage}
\begin{abstract}
We use high-resolution hydrodynamical simulations of large-scale
structure formation to study the imprints of the local superclusters
onto the full-sky Sunyaev-Zel'dovich (SZ) signals. Following 
\cite{Mathis:2002}, the initial conditions have been statistically 
constrained to reproduce the density field within a sphere of 110 Mpc
around the Milky Way, as observed in the IRAS 1.2-Jy all-sky redshift
survey.  As a result, the positions and masses of prominent galaxy
clusters and superclusters in our simulations coincide closely with
their real counterparts in the local universe.  We present the results
of two different runs, one with adiabatic gas physics only, and one
also including cooling, star formation and feedback.  By analysing the
full-sky maps for the thermal and kinetic SZ signals extracted from
these simulations, we find that for multipoles with $\ell<100$ the
power spectrum is dominated by the prominent local superclusters, and
its amplitude at these scales is a factor of two higher than that
obtained from unconstrained simulations; at lower multipoles
($\ell<20$) this factor can even reach one order of magnitude.  We
check the influence of the SZ effect from local superclusters on the
cosmic microwave background (CMB) power spectrum at small multipoles
and find it negligible and with no signs of quadrupole-octopole
alignment.  However, performing simulations of the CMB radiation
including the experimental noise at the frequencies which will be
observed by the Planck satellite, we find results suggesting that an
estimate of the SZ power spectrum at large scales can be extracted.
\end{abstract}


\section{Introduction}

It is well known that high-accuracy measurements of the anisotropies
of the cosmic microwave background (CMB) are extremely important for
cosmology. By analysing their angular power spectrum, $C(\ell)$, it is
in fact possible to obtain tight constraints on the mechanisms from
which the primordial fluctuations originate and on the main
cosmological parameters (or combinations of them) like the matter
density, the baryonic density and the cosmological constant
contribution. For this reason, the astronomical community has been
doing a big effort over the last years in order to improve the quality
of the CMB data, both in terms of accuracy and resolution.

The best available CMB measurements up to now were provided by the
{\it Wilkinson Microwave Anisotropy Probe} satellite
\citep[WMAP;][]{BE03.1}.  The data obtained in its first year of
activity gave strong support to the so-called concordance model,
i.e. a spatially flat universe dominated by dark energy and cold dark
matter, with nearly scale-invariant primordial fluctuations, almost
gaussianly distributed.  The WMAP result is confirming the general
picture emerging by the analysis of other different datasets, like
high-redshift supernovae Ia, weak lensing, galaxy survey,
Lyman-$\alpha$ forest and galaxy clusters \citep[see discussion in][
and references therein]{SP03.1}.

The statistical analysis and the theoretical interpretation of CMB
data of so high quality necessarily require a good understanding of
all possible sources of contamination as well as the techniques to
separate them from the CMB signal.  The Sunyaev-Zel'dovich (SZ) effect
\citep{sz}, i.e. the Compton scatter of the cold CMB photons by the
hot electrons mainly located in the intracluster plasma, certainly
represents one of the most important secondary anisotropies, i.e.
temperature fluctuations generated along the light trajectory from the
last scattering surface to the observer. Different studies have been
devoted to the estimate of the SZ angular power spectrum which
originates from the whole galaxy cluster population \citep[see, e.g.,
][ and references therein]{cooray2004}. The results show that the SZ
signal starts to dominate the cosmological one only on very small
angular scales ($\ell>2000$), while at the scales covered by WMAP is
approximately 3 orders of magnitude smaller.

Even if its amplitude appears relatively small, the SZ signal could be
nevertheless significant at the sky positions corresponding to nearby
galaxy clusters.  This has motivated the research of SZ imprints in
the first-year WMAP temperature maps. This approach, mainly based on
the cross-correlation of the CMB maps with the observed distribution
of clusters and superclusters, as extracted in both optical and X-ray
surveys, led to different claims for a positive detection at small
angular scales \citep{myers2004,2004ApJ...613L..89H,fosalba2004,afshordi2004b},
but only upper limits for the detection of the SZ signal exist at
larger scales at present
\citep{diego2003,hern2004b,hirata2004,huffenberger2004,hansen}.
In this sense the situation will largely improve in the near future
thanks to the capabilities of the {\it Planck} satellite, whose launch
is planned for 2007. In fact the range of frequencies covered by its
receivers will be very large, and the noise level is expected to be
very low. This will allow to extract the SZ signal for at least 10,000
objects out to redshfit $z\approx 1$ \citep[see,
e.g.][]{bartelmann2001}.

The CMB data from the WMAP satellite exhibit several anomalies at the
lowest multipoles, among them the abnormally low value of the
quadrupole amplitude \citep[see, however,][]{efsta2004} and the fact
that the directions of the quadrupole and octopole are unusually well
aligned \citep[see, e.g.,][]{gazta2003,tegmark2}.  There have been
some speculation about the possibility that the SZ signal from
structures in the local universe (the so-called local supercluster)
could be the cause of the low quadrupole \citep{brasiliani}. Further,
the quadrupole and octopole directions happen to be pointing toward
the Virgo cluster \citep{tegmark2}, which opens the possibility of
contamination from the local universe.

The goal of this paper is to investigate the importance of the thermal
and kinetic SZ effects produced by the local superclusters by
comparing it with respect to the cosmological signal and to the
contribution of more distant clusters.  At this aim we will use the
results of high-resolution hydrodynamical simulations realized
starting from initial conditions \citep{Mathis:2002} which are
constrained to reproduce the density field recovered by the IRAS
1.2-Jy all-sky redshift survey.  Using full-sky maps extracted from
these numerical simulations, we will discuss the statistical
properties of low multipoles in terms of amplitude and
alignment. Moreover, taking advantage of the different set of physical
processes treated in the simulations we can assess the uncertainties
related to the modelization. Finally we will perform simulations to
discuss if the local SZ signal can be detected exploiting the
capabilities of the Planck satellite. 

The plan of the paper is as follows. In Section 2 we describe the
general characteristics of the constrained hydrodynamical simulations
of the local universe used in the following analysis. Section 3
presents our method to produce full-sky maps of the thermal and
kinetic SZ effects.  Section 4 is devoted to the results of our
analysis. In particular we show the statistical properties of the maps
in terms of angular power spectra and alignment of lower multipoles
and we discuss the possibility of detecting the local SZ signal with
future experiments, like the upcoming Planck satellite. Finally we
summarize our findings and draw the main conclusions in the final
Section 5.

\section{The constrained hydrodynamical simulations}

The results presented in this paper have been obtained by using the
final output of two different cosmological hydrodynamical simulations
of the local universe obtained starting from the same initial
conditions but following a different set of physical processes.  In
more detail we used initial conditions similar to those adopted by
\citet{Mathis:2002} in their study (based on a pure N-body simulation)
of structure formation in the local universe.  The galaxy distribution
in the IRAS 1.2-Jy galaxy survey is first gaussianly smoothed on a
scale of 7 Mpc and then linearly evolved back in time up to $z=50$
following the method proposed by \cite{Kolatt:1996}. The resulting
field is then used as a Gaussian constraint \citep{Hoffman1991} for an
otherwise random realization of a flat $\Lambda$CDM model, for which
we assume a present matter density parameter $\Omega_{0m}=0.3$, a
Hubble constant $H_0=70$ km/s/Mpc and a r.m.s. density fluctuation
$\sigma_8=0.9$.  The volume that is constrained by the observational
data covers a sphere of radius $\sim 110$ Mpc, centred on the Milky
Way. This region is sampled with more than 50 million 
high-resolution dark matter particles
and is embedded in a periodic box of $\sim 343$ Mpc on a side. The
region outside the constrained volume is filled with nearly 7
million low-resolution dark matter particles, allowing a good 
coverage of long-range gravitational tidal forces.

The statistical analysis made by \citet{Mathis:2002} demonstrated that
the evolved state of these initial conditions provides a good match to
the large-scale structure observed in the local universe.  Using
semi-analytic models of galaxy formation built on top of merging
history trees extracted from the dark matter distribution in the
simulation, they showed that the density and velocity maps obtained
from synthetic mock galaxy catalogues have characteristics very
similar to their observational counterparts.  Moreover many of the
most prominent nearby galaxy clusters like Virgo, Coma, Pisces-Perseus
and Hydra-Centaurus, can be identified directly with haloes in the
simulation, with a good agreement for sky positions and virial masses.
In fact the positions of all identified objects differ less than the
smoothing radius adopted in the initial conditions (7 Mpc), with the
only exception of Centaurus, which is displaced by 9.6 Mpc. The
agreement for the masses is within a factor of 2 which can be
considered acceptable if one considers the uncertainties in the
inferred observational masses, mainly when velocity dispersion is
used.

Unlike in the original simulation made by \citet{Mathis:2002}, where
only the dark matter component is evolved, here we want to follow also
the gas distribution. For this reason we extended the initial
conditions by splitting the original high-resolution dark matter
particles into gas and dark matter particles having masses of $0.48
\times 10^9\; M_\odot$ and $3.1 \times 10^9\; M_\odot$, respectively;
this corresponds to a cosmological baryon fraction of 13 per
cent.  The total number of particles within the simulation is then
slightly more than 108 million and the most massive clusters will be
resolved by almost one million particles.

Our runs have been carried out with {\small GADGET-2}
\citep{springel2005}, a new version of the parallel Tree-SPH
simulation code {\small GADGET}
\citep{SP01.1}.  The code uses an entropy-conserving formulation of
SPH \citep{2002MNRAS.333..649S}, and allows a treatment of radiative
cooling, heating by a UV background, and star formation and feedback
processes.  The latter is based on a sub-resolution model for the
multiphase structure of the interstellar medium
\citep{2003MNRAS.339..289S}. The code can also follow the pattern of
metal production from the past history of cosmic star formation
\citep{2004MNRAS.349L..19T}.  This is done by computing the contributions 
from both Type-II and Type-Ia supernovae and energy feedback and
metals are released gradually in time, accordingly to the appropriate
lifetimes of the different stellar populations. This treatment also
includes in a self-consistent way the dependence of the gas cooling on
the local metallicity.

As said, the results presented in this work are based on two different
simulations starting from the same initial conditions.  The first one,
hereafter called {\it gas}, includes only non-radiative
(i.e. adiabatic) hydrodynamics and has been originally analyzed by
\citet{2005JCAP...01..009D} to study the propagation of cosmic rays in
the local universe.  \citet{hansen} used the gas properties extracted
from this simulation to infer a prediction of the SZ effect from
diffuse hot gas in the local universe. In that paper, the PSCz
catalogue was used to map the matter density whereas the simulation
was used only to assign a temperature for the gas with a given
density. The SZ map realized in this way was cross-correlated to the
WMAP data and upper limits were found. The second run, hereafter
called {\it csf}, exploits all the capabilities of the present version
of the GADGET code, including cooling, star formation, feedback and
metallicity; its outputs will be also used to study the detectability
of the diffuse, warm intergalactic medium with future X-ray
experiments (Kawahara et al. in preparation). 

The feedback scheme for the {\it csf} run assumes a Salpeter IMF
\citep{1955ApJ...121..161S} and its parameters have been fixed  to get 
a wind velocity of $\approx 480$ km/s.  In a typical massive cluster
the SNe (II and Ia) add to the ICM as feedback $\approx 2$keV per
particle in an Hubble time (assuming a cosmological mixture of H and
He); $\approx 25$ per cent of this energy goes into winds.  Note that
these values can be considered an upper limit because a part of the
ICM affected by the star processes could be at the present time out of
the cluster virial radius.  Moreover such feedback mechanisms cannot
avoid the overcooling problem, commonly found in cosmological
simulations: the corresponding overproduction of stars, acting mainly
in the central regions, tends to amplify by a factor of $\approx 2$
the effects of the star population.  Notice that the metal-dependence
of the cooling function does not significantly change the global
feedback properties.  Another signature that the feedback scheme is
not enough efficient is the large amount of metals which is still
locked inside the star particles: as a consequence, the resulting ICM
metallicity is low, even if still compatible with observed values.  A
more detailed discussion of cluster properties and metal distribution
within the ICM as resulting in simulations including the metal
enrichment feedback scheme can be found in
\citet{2004MNRAS.349L..19T}.

The gravitational force resolution (i.e. the comoving softening
length) of both simulations has been fixed to be 14 kpc
(Plummer-equivalent), which is comparable to the interparticle
separation reached by the SPH particles in the dense centres of our
simulated galaxy clusters.

In Table \ref{tab:tab0} we report the main characteristics of the most
prominent clusters in our simulations. In particular, we computed the
(mass-weighted) temperature within one tenth of the virial radius, and
the central Compton-$y$ parameter.  In agreement with previous results
in the literature \citep[see, e.g., ][]{tornatore2003}, we find that
the temperatures for a given object in the ${\it csf}$ simulation are
always larger than in the ${\it gas}$ simulation by a factor of
approximately 15-20 per cent.  Including radiative cooling causes in
fact a lack of pressure support, with a subsequent heating of the
infalling gas by adiabatic compression.  The opposite trend is present
for the central Compton-$y$ parameter, which can be larger by a factor
up to 2 in the ${\it gas}$ simulation than in the ${\it csf}$ one
\citep[see also][]{2002ApJ...579...16W,dasilva2004,motl2005}.  For
completeness, we also list the available observational estimates,
which are in rough agreement.  Notice that only for Coma we averaged
in the simulation the value for Compton-$y$ within an area of 450 kpc
to be compatible with the beam size of the observational data
\citep{2003ApJ...598L..75B}. Finally we would like to remark that the
observed temperatures are not directly comparable with the simulation
results, because they are extracted from X-ray spectroscopic data: as
shown by \cite{mazzotta:2004} \citep[see
also][]{rasia:2005,vikhlinin:2005}, there could be a significant bias
produced by the complexity of the thermal structure. For
completeness we added in the table the values for the
spectroscopic-like temperature $T^{\rm SL}$, which
approximates the spectroscopic temperature better than few per cent
\citep{mazzotta:2004}.

\begin{table*}
   \begin{center} \caption{Comparison of the properties of the 6 most
   prominent clusters, listed in Column 1. The observed cluster
   temperatures $T_{\rm obs}$ (Column 2) are taken from
   \citet{1999ApJ...517..627M}, with the exception of the data for
   Virgo \citep{2003MNRAS.340..989S} and A3627
   \citep{2002A&A...383..773I}. Columns 3 and 5 refer to the
   mass-weighted temperature within $0.1\times R_{\rm vir}$, as
   computed in the {\it gas} and {\it csf} simulations ($T_{\rm
   gas}^{\rm MW}$ and $T_{\rm csf}^{\rm MW}$, respectively).
   Columns 4 and 6 refer to the spectroscopic-like temperature
   \citep{mazzotta:2004} within $0.1\times R_{\rm vir}$, as computed
   in the {\it gas} and {\it csf} simulations ($T_{\rm gas}^{\rm SL}$
   and $T_{\rm csf}^{\rm SL}$, respectively). In Column 7 we report
   the value for the Compton-$y$ parameter, as obtained for Coma by
   \citet{2003ApJ...598L..75B}.  Columns 8 and 9 refer to the values
   for $y$ computed in the {\it gas} and {\it csf} simulations
   ($y_{\rm gas}$ and $y_{\rm csf}$, respectively).  For Coma we
   averaged the Compton-$y$ within an area of 450 kpc to be compatible
   with the beam size of the observational data
   \citep{2003ApJ...598L..75B}.}

   \begin{tabular}{l|c|c|c|c|c|c|c|c|c|}
   Cluster   & $T_{\rm obs} [\rm kev]$     
             & $T_{\rm gas}^{\rm MW} [\rm kev]$     
             & $T_{\rm gas}^{\rm SL} [\rm kev]$     
             & $T_{\rm csf}^{\rm MW} [\rm kev] $ 
             & $T_{\rm csf}^{\rm SL} [\rm kev] $ 
             & $y_{\rm obs} [10^{-5}]$ 
             & $y_{\rm gas} [10^{-5}]$ 
             & $y_{\rm csf} [10^{-5}]$\\
   Coma      & $8.21^{+0.16}_{-0.16}$ & 6.1 & 5.8 & 7.3 & 7.3 & $9.6^{+1.1}_{-1.1}$ 
& 9.2 & 6.4 \\
   Virgo     & $2.55^{+0.07}_{-0.06}$ & 3.5 & 3.0 & 4.1 & 3.5 & & 4.8 & 3.1 \\
   Centaurus & $3.54^{+0.08}_{-0.08}$ & 3.8 & 3.6 & 4.6 & 4.5 & & 6.7 & 4.2 \\
   Hydra     & $3.10^{+0.11}_{-0.11}$ & 3.2 & 3.2 & 4.5 & 4.6 & & 9.3 & 4.2 \\
   Perseus   & $6.33^{+0.21}_{-0.18}$ & 5.8 & 5.2 & 6.1 & 5.8 & & 9.5 & 6.8 \\
   A3627     & $5.62^{+0.12}_{-0.11}$ & 3.7 & 3.6 & 4.4 & 4.4 & & 6.6 & 4.2 \\
   \end{tabular}
   \label{tab:tab0} \end{center}
\end{table*}

\section{Full sky map making}

Hot electrons, which are present in the intracluster medium, scatter
by inverse Compton the cold photons of the cosmic microwave background
(CMB) radiation and re-distribute them towards higher frequencies.
The result is the so-called SZ effect \citep{sz} which originates a
temperature decrement below 217~GHz, and an increment above.

Given a direction $\vec{\theta}$, the change of the CMB temperature
$T_{\rm CMB}$ produced by the thermal SZ effect is given by
\be
\frac {\Delta T_{\rm CMB}} { T_{\rm CMB}} = y(\vec{\theta}) g(x)\ ,
\ee
where $g(x)= x\cdot {\rm{coth}}(x/2)-4$ and $x\equiv h \nu /k T_{\rm
CMB}$.

The Compton-$y$ parameter is related to the three-dimensional thermal
electron density, $n_e$, and to the electron temperature, $T$, by
\be
y(\vec{\theta})= {{k\sigma_{_T}}\over{m_e c^2}} \int \dd l
\ n_e(\vec{\theta},l)\ T(\vec{\theta},l)\ ,
\label{eq:y_theta}
\ee
$\sigma_{_T}$ being the Thomson scattering cross section. 

The kinetic SZ effect is produced by the motion of the intracluster
gas with respect to the CMB. The corresponding change of the CMB
temperature can be written as
\be
\frac {\Delta T_{\rm CMB}} { T_{\rm CMB}} = - w(\vec{\theta})\ ,
\ee
with
\be
w(\vec{\theta})=\frac{\sigma_{_T}}{c} \int \dd l
\ n_e(\vec{\theta},l)\ v_r(\vec{\theta},l)\ ,
\label{eq:w_theta}
\ee
where $v_r$ is the radial component of the cluster velocity.

The line of sight integrals which are present in the definitions of
$y$ (eq.~\ref{eq:y_theta}) and $w$ (eq.~\ref{eq:w_theta}) can be
computed in the simulations by exploiting the SPH kernel. Here we
employ the spline kernel defined as \citep{1985A&A...149..135M}
\be
   W(r/h)=\frac{\pi}{h^3}\left\{\begin{array}{ll}
      1-\frac{3}{2}(r/h)^2+\frac{3}{4}(r/h)^3 & 0 \le (r/h) \le 1 \\
      \frac{1}{4}(2-r/h)^3              & 1 \le (r/h) \le 2 \\
      0                               & 2 \le (r/h) \\
   \end{array} \right. , \label{sKern}
\ee
where $h$ is the so-called SPH smoothing length and $r$ represents the
impact parameter.  In this way eq.(\ref{eq:y_theta}) becomes
\be
  y_{\rm int}=\int
\sum_j \frac{m_j}{\rho_j}
          y_j W(d_j(r)/h_j)\dd r\ ,
\ee
where $m_j$ and $\rho_j$ are the mass and density of the $j$-th gas
particle, $y_j$ is the corresponding Compton $y$-parameter and
$d_j(r)$ is the projected distance with respect to the position $r$
along the line of sight. A similar relation can be written for $w_{\rm
int}$.

In principle the sum has to be done over all particles, but in the
case of a compact kernel where $W(r/h)$ gets zero at large $r/h$ (as
the one we use) it can be restricted to those particles having a distance
from the line of sight smaller than twice their smoothing length.

In practice, however, one would like to obtain the value averaged over
all lines of sight crossing the pixel, instead of the value along the
line of sight crossing the centre of the pixel only.  This can become
a problem when the projected size of the structure producing the
signal gets smaller than the pixel size.  Therefore one usually
applies the so called {\it gather approximation}, where, given a
pixel, all particles - whose projections either overlap or completely
fall inside the pixel - are taken into account.  In this case the
value of the Compton-$y$ parameter can be written as the sum of the
contributions of all relevant particles as:
\be
  y_{\rm pix}=\sum_j \frac{m_j}{\rho_j}
   y_j N_j \frac{A_{\rm pix}}{A_j} W_{\rm int}(\tilde{d_j}/h_j)\ ,
\ee
where $\tilde{d_j}$ is the projected distance between the particle and
the pixel centre and $W_{\rm int}$ is the integrated kernel.  The area
$A_j$ associated to each gas particle can be approximated as the
square of the cubic root of the corresponding volume,
i.e. $A_j=(m_j/\rho_j)^{2/3}$.  The normalization factor $N_j$ is
given by
\be
  \sum_{\rm pix} N_j A_{\rm pix} W_{\rm int}(\tilde{d_j}/h_j)
   = A_j
\ee
to ensure the conservation of the quantity $y$ when distributed over
more than one pixel.

The method needs two further corrections. First, for pixels which are
only partially overlapped by the projected SPH smoothing kernel, we
approximate the associated area $\hat{A}_{\rm pix}$ as
\be
  \hat{A}_{\rm pix} = \sqrt{A_{\rm pix}}
    \times(\tilde{d_j}+0.5\sqrt{A_{\rm pix}}-h_j)\ .
\ee
To compute the contribution to such a pixel, the weight of the
integrated kernel $W_{\rm int}$ is then corrected by a factor
$\hat{A}_{\rm pix}/A_{\rm pix}$. This must be taken also into account
when normalizing the integrated kernel. We checked that this
correction leads to very small changes in the majority of the pixels,
but its effect can become substantial in the high-signal regions
corresponding to the cluster cores, specially when the particles are
distributed over a small number of cells.

Second, when a particle contributes to one pixel only, we fix the
integrated kernel $W_{\rm int}$ to be unity.  This is important when
the area associated to such particle is much smaller than the pixel
area and ensures that the value corresponding to the particle is
completely given to this pixel: in this case the normalization factor
$N_j$ can become significantly smaller than unity to ensure the
conservation of the quantities.  Finally we notice that, since the SZ
effect is independent of distance, no distance factor has to be
included into the previous equations.

\begin{figure}
  \begin{center}
    \includegraphics[width=0.45\textwidth]{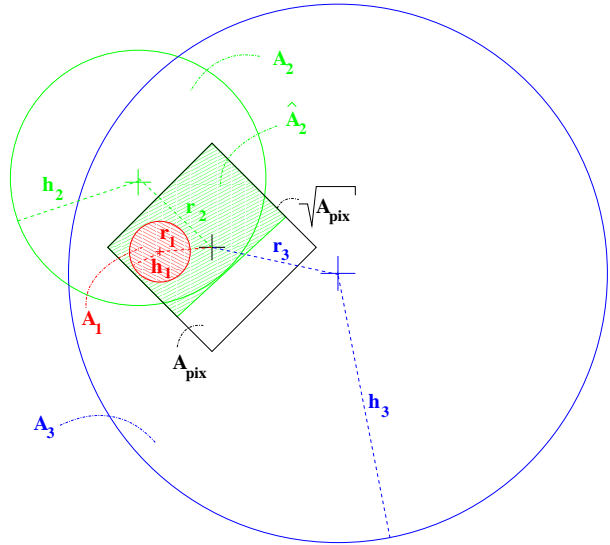}
  \end{center}
  \caption{
Sketch of three different geometrical situations for particles
contributing to one pixel (represented by the black square).  Particle
1 (in red), having a smoothing length $h_1$ and being at a distance
$r_1$ from the pixel centre, falls completely inside the pixel, but
does not overlap the centre.  Particle 2 (in green) covers only
partially the pixel. For particle 3 (in blue) the whole pixel lies
within its projected radius.
  \label{fig:sketch}  }
\end{figure}

Figure \ref{fig:sketch} exemplifies the three different situations
previously discussed by showing how the gas particles can contribute
to one pixel (here represented by the square).  In such cases, the
weight contributed by particle 1 (shown in red) is $A_1/A_{\rm pix}$,
the weight for particle 2 (in green) is $W_{\rm int}(r_2/h_2)
\hat{A}_2/A_2$ and the weight for particle 3 (in blue) is $W_{\rm
int}(r_2/h_2) A_{\rm pix}/A_3$. Note that for particles 2 and 3 the
weight needs to be normalised to unity when summed over all pixels to
which the particles are contributing.

Our maps have been constructed by applying the previous technique
directly to the \hp~ representation of the full sky \citep{healpix}.
We fix the {\it nside} parameter to be 1024, which results in an
all-sky representation with $\approx 1.2\times10^7$ pixels.  In order
to avoid spurious effects the line of sight integral has been
performed between a minimum and maximum distance ($r_{\rm min}$ and
$r_{\rm max}$, respectively).  In particular we fix $r_{\rm min}=5$
Mpc to be sure that the observer is outside the SPH smoothing radius
of all particles, and $r_{\rm max}=110$ Mpc, which represents a
conservative limit of our high-resolution region.

\section{Results}

\begin{figure*}
  \begin{center}
    \includegraphics[width=\textwidth]{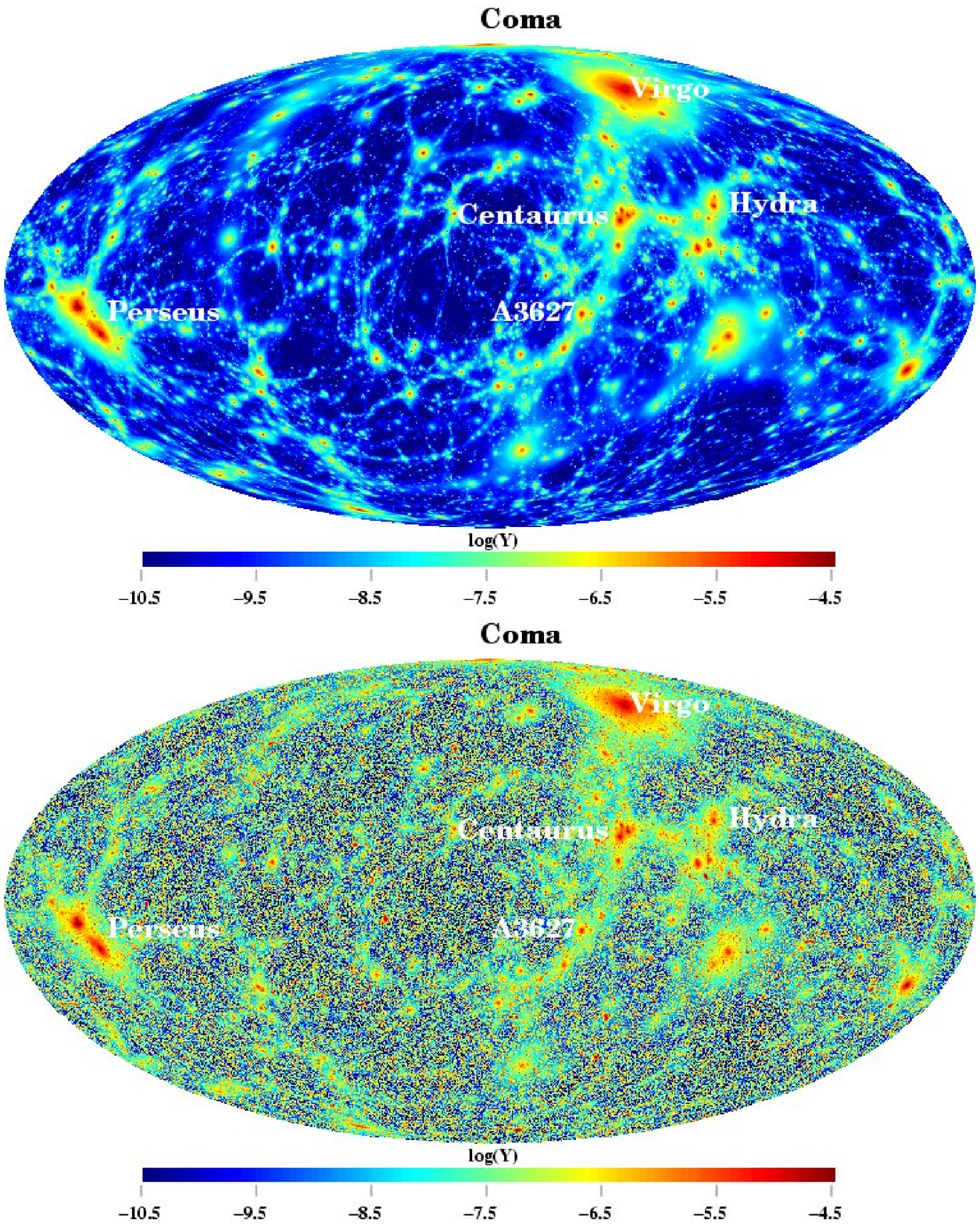}
  \end{center}
  \caption{
 Full-sky map of Compton-$y$ parameter in galactic coordinates.  The
 upper panel shows the results from the {\it gas} simulation, covering
 the local universe up to 110 Mpc from the Milky Way, while in the
 lower panel we added the contribution from more distant objects, as
 estimated by Schaefer et al. (2004a).  The position of the most
 prominent structures is indicated in the maps.  }
\label{fig:fullsky1}
\end{figure*}

\begin{figure*}
  \begin{center}
    \includegraphics[width=\textwidth]{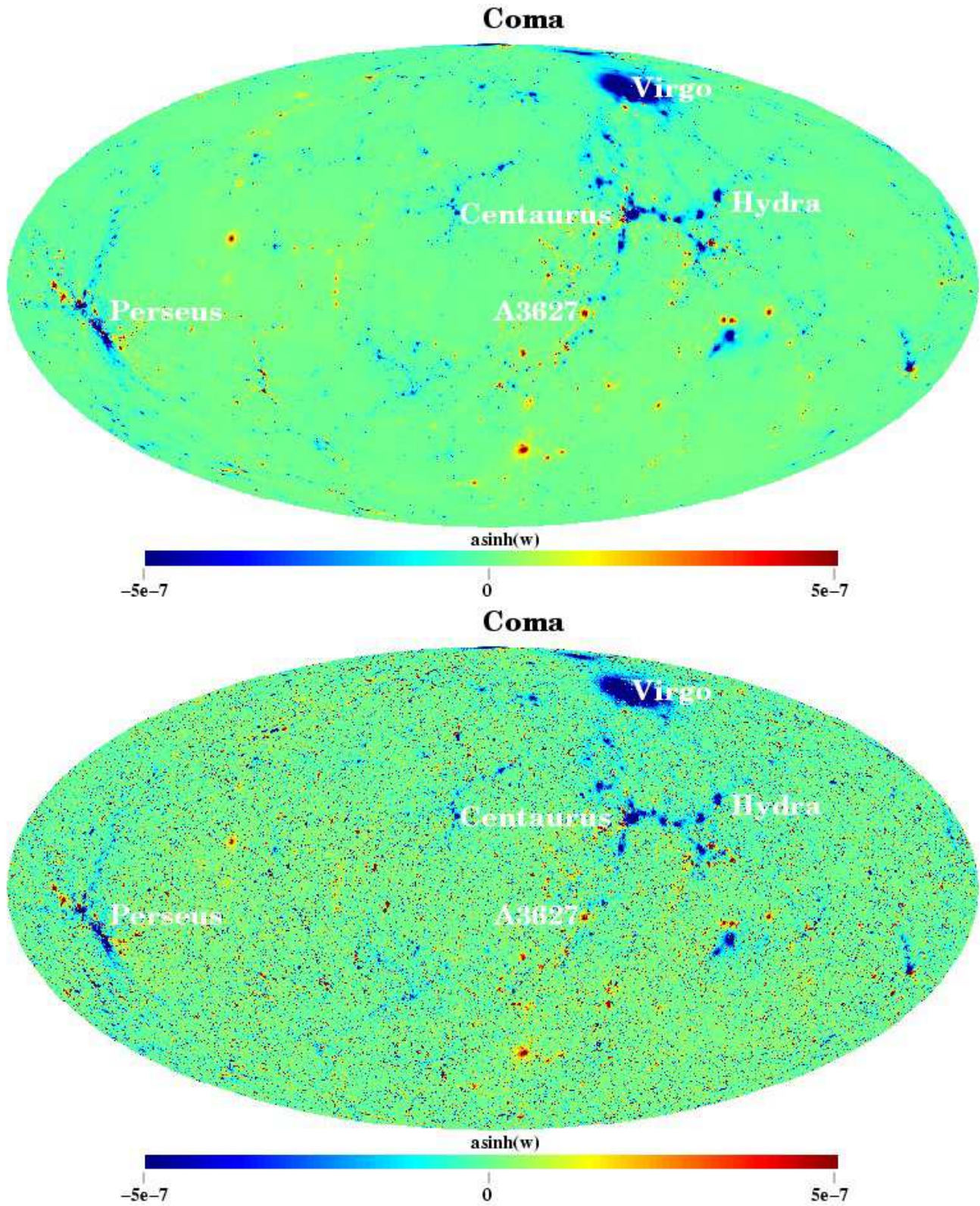}
  \end{center}
  \caption{
  The same as Fig.\ref{fig:fullsky1}, but for the kinetic SZ effect,
  shown by using ${\rm asinh}(w)$ to better display positive and
  negative values.}
\label{fig:fullsky2}  
\end{figure*}

\subsection{Full-sky maps}

\begin{figure*}
    \includegraphics[width=8cm]{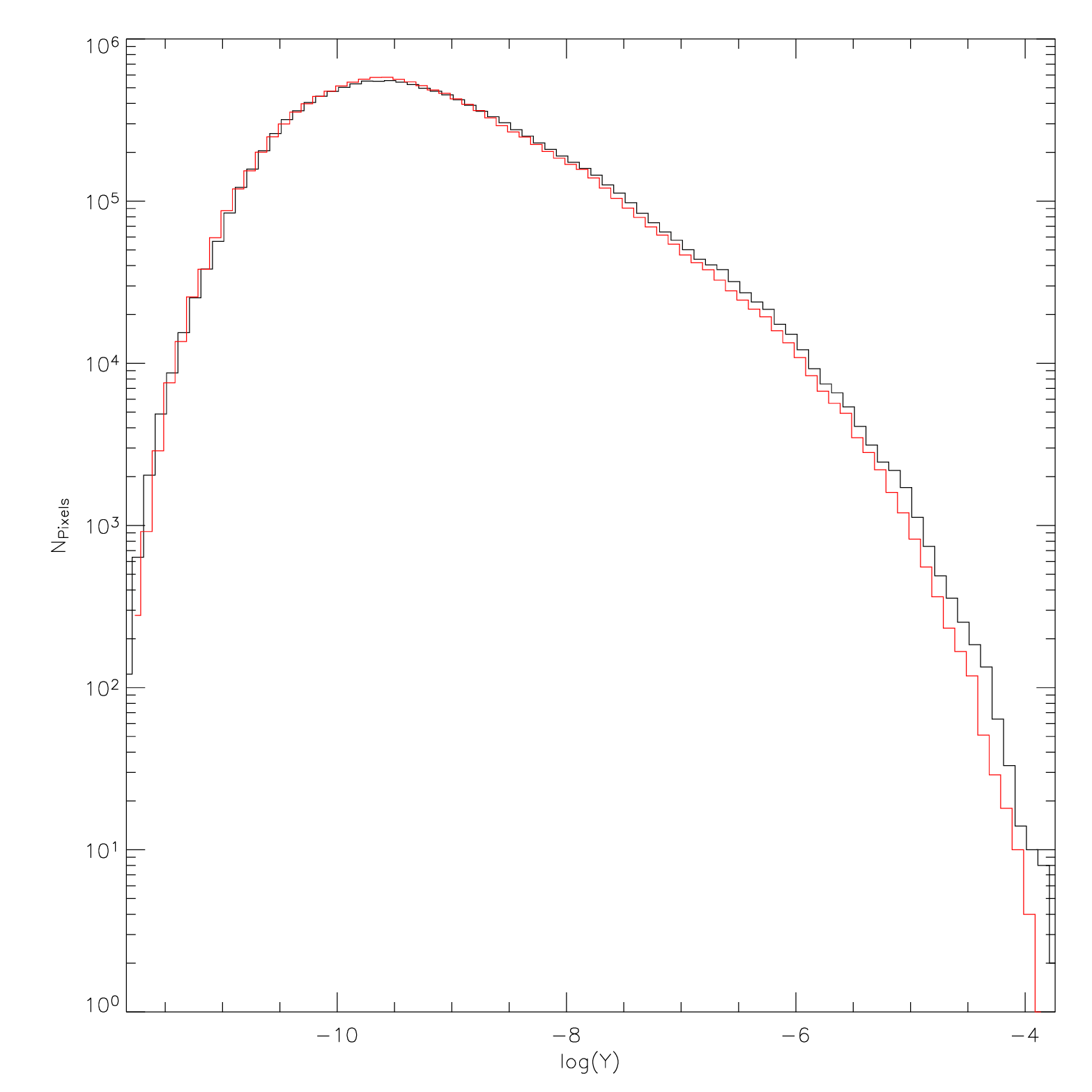}
    \includegraphics[width=8cm]{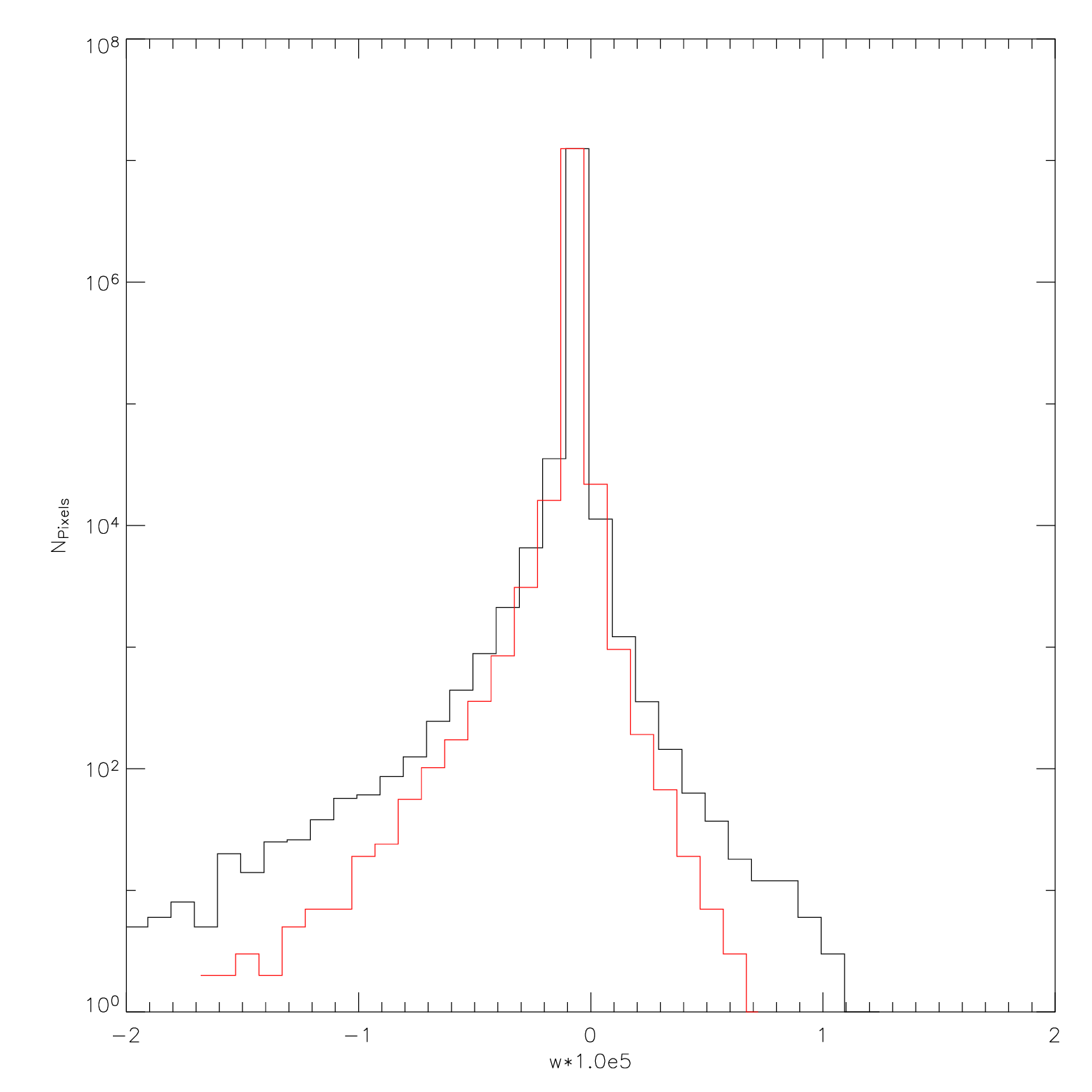}
\caption{
Comparison between the SZ signals as extracted from the {\it gas} and
{\it csf} simulations (black and red lines, respectively).  The number
of pixels having a given value for the Compton-$y$ parameter and the
kinetic parameter $w$ are shown in the left and right panels,
respectively.}
\label{fig:histpixel}
\end{figure*}

In Fig. \ref{fig:fullsky1} we show the results of the application of
the previous method to obtain a full-sky map of the Compton-$y$
parameter in supergalactic coordinates.  The upper panel corresponds
to the signal resulting from our {\it gas} simulation, covering the
local universe up to $110$ Mpc.  It is possible to recognize the most
prominent features: Perseus, Virgo, Centaurus, Hydra, A3627 and Coma
(very close to the northern pole of the map).  Also the presence of
the filamentary structure connecting the largest clusters is still
evident.  In the lower panel we show the same map where we add the
contribution coming from more distant objects. This has been computed
by exploiting the results of Schaefer et al. (2004a,b), who build SZ
signal maps by using the {\it Hubble volume simulation}
\citep{2000MNRAS.319..209C,2001MNRAS.321..372J} and suitably
inserting the outputs of different hydrodynamic re-simulations of
single clusters. Notice that the physical processes considered in
these re-simulations, i.e. non-radiative hydrodynamics, are the same
included in our {\it gas} simulation, so no biases are introduced in
combining the results.  With respect to the \cite{schaefer2004a} work
we add only the objects having a distance from the observer larger
than 110 Mpc, because the closer ones are already represented in our
simulation.  The resulting map appears more patchy, having a
significant contribution coming from a large number of distant
clusters.

The corresponding maps for the kinetic SZ signal $w$ are shown in
Fig. \ref{fig:fullsky2}. Again the most important structures of the
local universe are evident in the upper map and continue to give the
dominant contribution in the map including distant objects, shown in
lower panel.

It is worth to notice that within the simulation we measure a velocity
at the observer' position of $v_x\approx -270$ km/s, $v_y\approx 420$
km/s and $v_z\approx -350$ km/s (assuming supergalactic coordinates).
This leads to a absolute velocity of $v\approx 610$ km/s, which is in
good agreement with the observed value \citep[see, e.g., the WMAP
analysis by][]{BE03.1}.  Again, the observer' velocity points toward a
direction which differs from the observed one less than 10 degrees:
this emphasizes once more that our simulations are giving a good
representation of the large-scale structure in our local universe.

By analyzing the maps obtained from the results of the {\it csf}
simulation (not shown here) we find only small differences with
respect to the previous results, mainly in correspondence of the
high-density regions, where the {\it cfs} simulation tends to give
smaller signals. This difference can be quantified by looking to the
pixel distribution as a function of $y$ and $w$. This is shown in
Fig.\ref{fig:histpixel}, where the contribution of the local universe
only is considered.  Concerning the thermal SZ effect (left panel),
the histograms are in general very similar, with only a slightly
higher frequency for higher values in the {\it gas} simulation. The
differences are more evident in the pixel distributions for the
kinetic parameter $w$: again high (absolute) values have higher
probability in the {\it gas} simulation than in the {\it csf}
one. 

Analyzing the gas bulk motions, we find no significant difference
in the {\it gas} and {\it csf} runs. Moreover the winds driven by the
feedback process are very efficiently stopped short after leaving the
star forming region and consequently are not contributing to the
kinetic SZ signal. Therefore the difference observed in
Fig. \ref{fig:histpixel} reflects the fact that the pressure in the
{\it csf} simulation is smaller than in the {\it gas} one. Again the
magnitude of the effect is in good agreement with previous findings 
\citep{2002ApJ...579...16W}.

Notice that the distribution for $w$ is clearly non-gaussian,
with a more extended tail toward negative values, produced by the
cluster motion inside the local universe. Since the spectral signature
of the kinetic SZ effect is the same of CMB, a perfect removal of this
contribution could be difficult.

It is also worth to notice that the mean value of the Compton-$y$
parameter computed over the whole sky is dominated by the background
map extracted from the {\it Hubble volume simulation}. For the total
map (Fig. \ref{fig:fullsky1}, lower panel) we get a mean value of
$3.3\times10^{-7}$, whereas we find a much smaller mean value
($4.1\times10^{-8}$) when considering only the map from the
constrained simulation ({\it gas} run). As discussed before, this
value slightly decreases when using the {\it cfs} simulation
($3.2\times10^{-8}$). Note that all these values are well below the
upper limit derived from COBE FIRAS data \citep{2003ApJ...594L..67F}.

\subsection{The angular power spectrum}

In order to quantify the amount of signal produced by the SZ effects,
we compute from our maps the angular power spectrum $C(\ell)$. In
particular we estimate the spectra in the Rayleigh-Jeans region, where
$\Delta T_{\rm CMB}/T_{\rm CMB}=-2y$.

\begin{figure}
    \includegraphics[width=8cm]{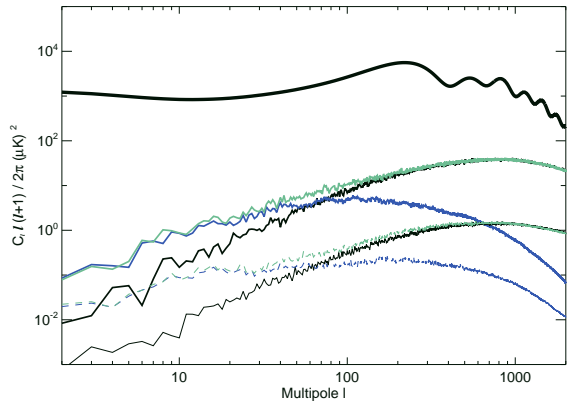}
  \caption{ 
The power spectra for the SZ effects.  Comparison between the
cosmological CMB power spectrum (upper black solid lined) and the SZ
signals obtained in the different simulations: {\it Hubble volume
simulation} (black lines), local constrained {\it gas} simulation
(blue lines); local constrained {\it gas} simulation combined with the
{\it Hubble volume simulation} for larger distances (green lines).
Upper and lower curves refer to the contributions from thermal and
kinetic SZ effects, respectively.  All spectra are taken in the
Rayleigh-Jeans region where $\Delta T_{\rm CMB}/T_{\rm CMB}=-2y$.  }
\label{fig:cl_all}
\end{figure}

The results are shown in Fig.\ref{fig:cl_all}, where we compare the
power spectrum of the primordial CMB radiation 
\cite[here represented by the upper black line corresponding to the
best fit of the WMAP data obtained by][]{SP03.1} to the signals
extracted from the numerical simulations here considered.  In
particular the plot shows $C(\ell)$ for the {\it gas} simulation and
distinguishes the contributions coming from the local universe from
the one produced by more distant galaxy clusters. Both the thermal and
kinetic SZ effects are considered.  The results confirm that for
$\ell<2000$ the cosmological signal is dominating the SZ one, being
approximately 3-4 orders of magnitude larger.  Moreover the kinetic
signal is always smaller by a factor of 10-20 than that produced by
the thermal SZ effect. By comparing the results of our simulation with
the estimates obtained by \cite{schaefer2004a}, who populated the dark
matter only {\it Hubble volume simulation}
\citep{2000MNRAS.319..209C,2001MNRAS.321..372J} with individual,
adiabatic cluster simulations, we notice that the SZ effect from the
local universe (objects with distance smaller than 110 Mpc from the
Milky Way) is the most important contribution up to scales
corresponding to $\ell \approx 100$. This is due to the larger angular
size of the local supercluster structure. On the contrary, the signal
at larger multipoles ($\ell>200$) seems to be dominated by the more
distant sources, which are lying outside of our simulated volume.

\begin{figure}
    \includegraphics[width=0.45\textwidth]{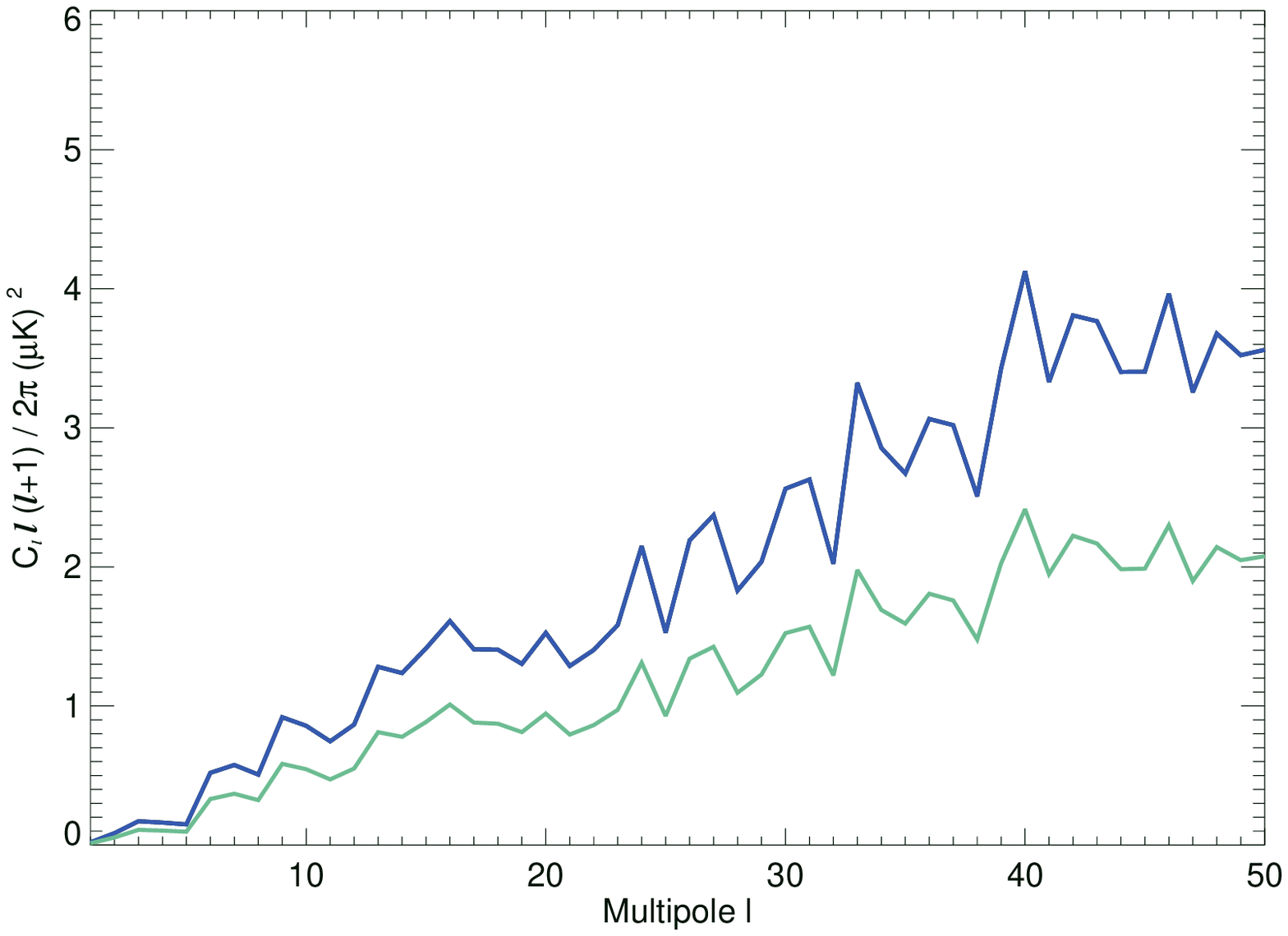}
    \includegraphics[width=0.45\textwidth]{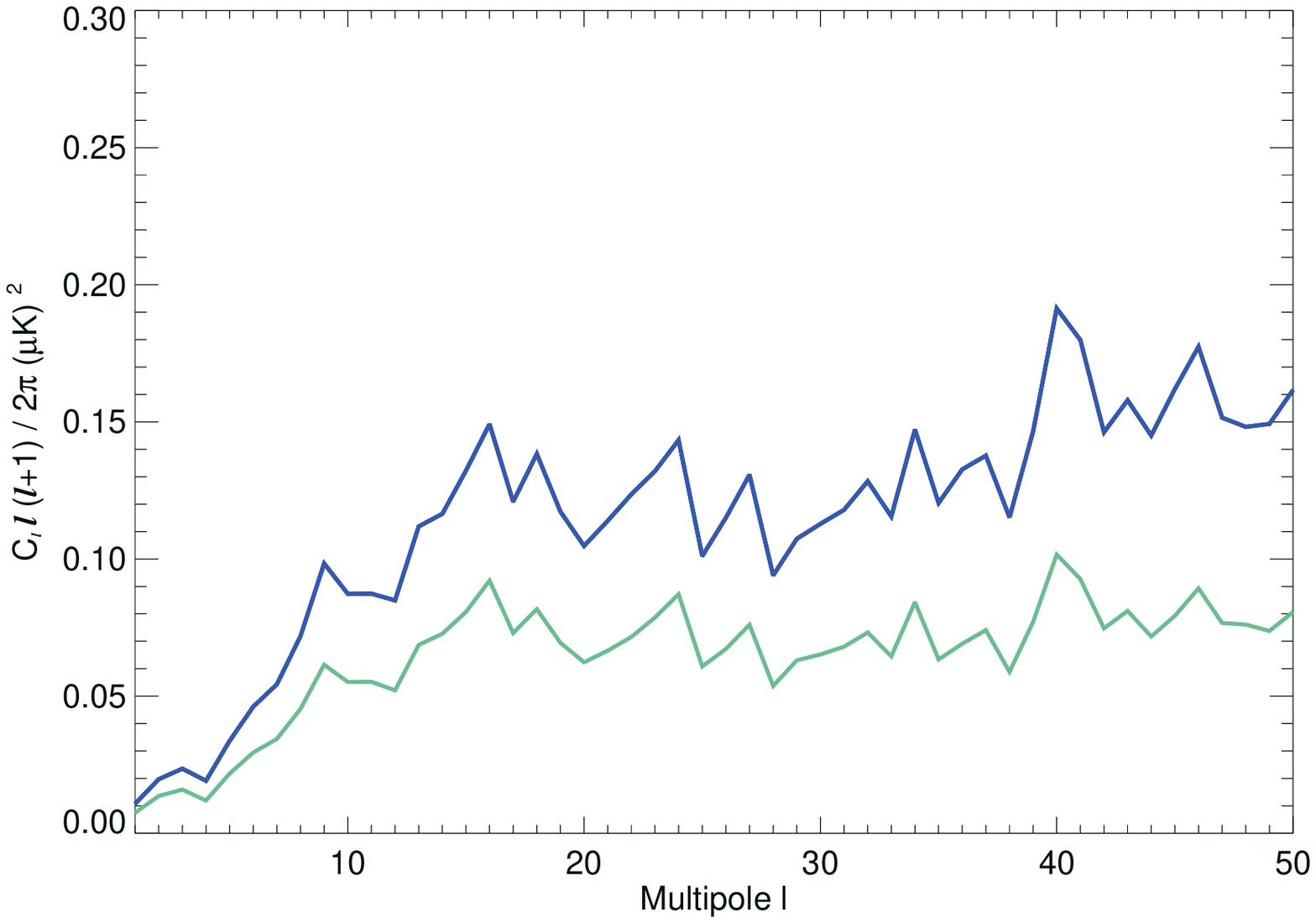}
   \caption{Effects on the power spectra of the inclusion of
   different physical processes in the constrained simulations of the
   local universe.  Blue and green curves show $C_l$ as extracted from
   the {\it gas} and {\it csf} simulations, respectively.  Upper and
   lower panels refer to the thermal and kinetic SZ effects,
   respectively. All spectra are taken in the Rayleigh-Jeans region
   where $\Delta T_{\rm CMB}/T_{\rm CMB}=-2y$.
\label{fig:cl_lowl}
}
\end{figure}

In order to quantify the contribution coming from the most
prominent clusters, we compute $C(\ell)$ from our maps by applying the
galactic cut and masking the regions where the 7 largest local galaxy
clusters are located. The resulting power spectrum (not shown here)
changes drastically for $\ell>50$ falling off much more rapidly, while
remains unchanged for $\ell<50$.  This is an expected result, because
the richest clusters are contributing to the power around 2-3 degrees,
whereas the total signal from all the rest is dominating the
large-scale power.

\begin{figure}
    \includegraphics[width=0.45\textwidth]{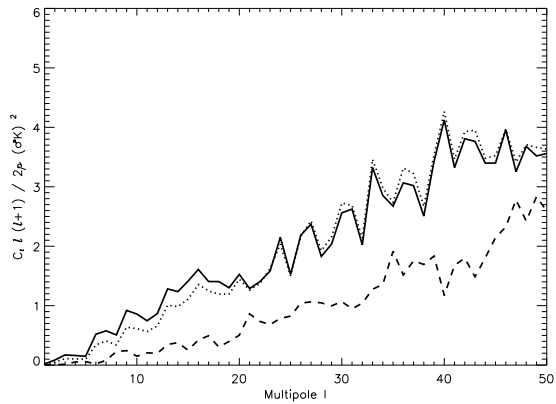} 
   \caption{The large-scale power spectrum of the SZ effect obtained from
    the constrained simulations including (solid line) and excluding
    (dotted line) the diffuse gas contribution. The dashed line shows
    the power spectrum of the {\it Hubble volume simulation}.
    \label{fig:nodiff}}
\end{figure}

We further checked the contribution from diffuse gas by masking
out in the maps the pixels with Compton-$y$ parameter smaller than a
given value ($y=10^{-6}$): in such a way only the signal from cluster
cores remains.  The resulting power spectrum is not strongly affected,
showing that the diffuse gas contributes minimally to the observed SZ
power. In Fig.~\ref{fig:nodiff} we show the power spectrum including
and excluding the diffuse gas contribution (solid and dotted lines,
respectively) along with the power spectrum extracted from the {\it
Hubble volume simulation} (dashed line).  Note that the {\it Hubble
volume simulation} gives a lower power spectrum at the largest scales
when compared to the constrained simulation, even when the diffuse gas
is removed from it.

Having two different constrained realizations of the local universe,
the first one with adiabatic physics only, the second one including
also cooling, star formation and feedback (the {\it gas} and {\it csf}
simulations, respectively), it is possible to discuss how the previous
results depend on the set of the included physical processes.  This is
done in Fig.\ref{fig:cl_lowl}, where we compare directly the power
spectra obtained from the two simulations.  We find a higher signal in
the adiabatic simulation, both for the thermal and kinetic SZ effects.
The amplitude of this difference is very similar in both cases, being
approximately a factor of 2.  This is in agreement with the results
obtained from individual cluster simulations
\citep[see, e.g.,][]{2002ApJ...579...16W}. 
Therefore we do not expect the signal to get stronger when including
additional physical processes and we can take the results obtained
from the {\it gas} simulation as a robust upper bound.

\subsection{Alignment of lower multipoles}

\begin{figure*}
\begin{center}
 \includegraphics[width=5cm,angle=90]{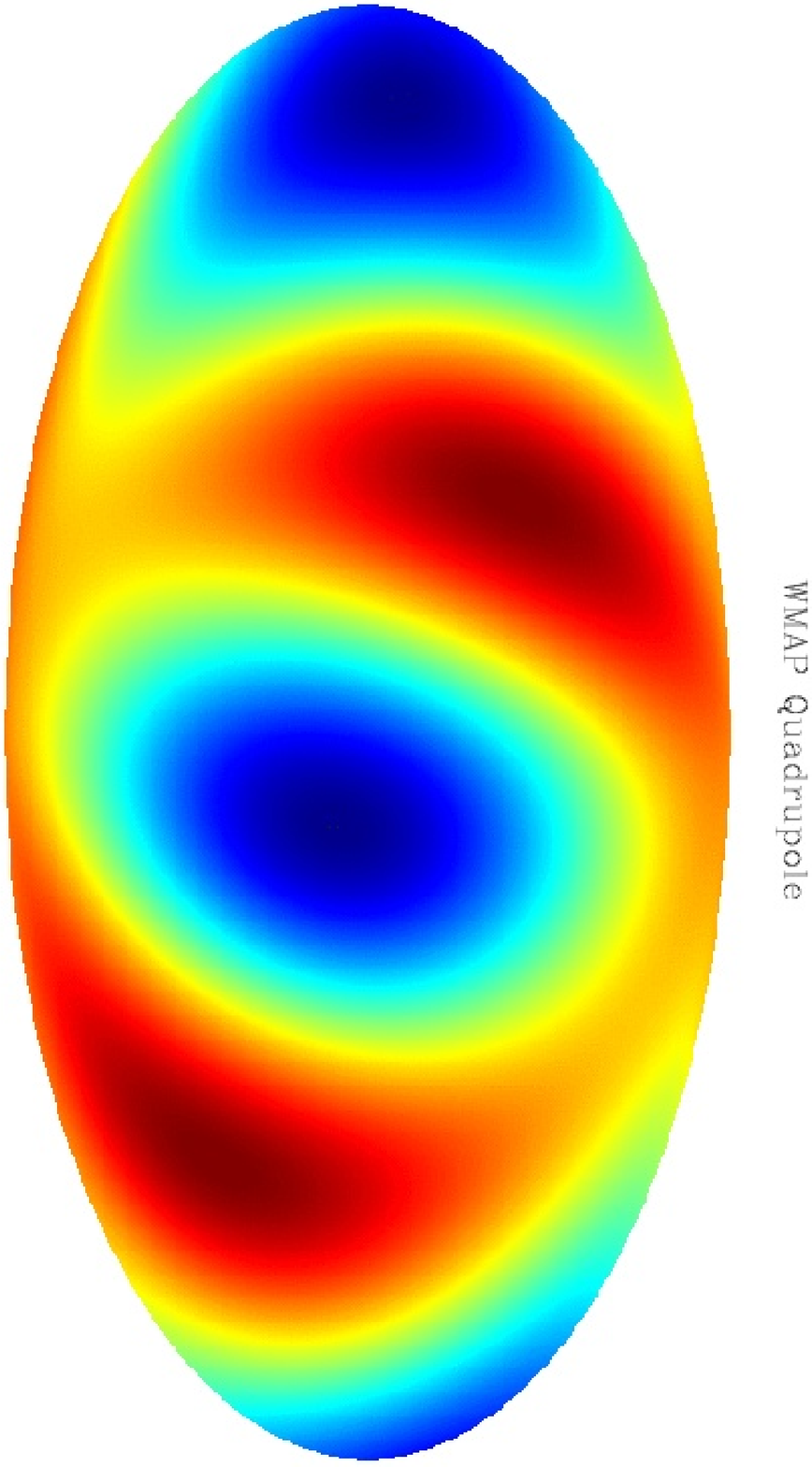}
 \includegraphics[width=5cm,angle=90]{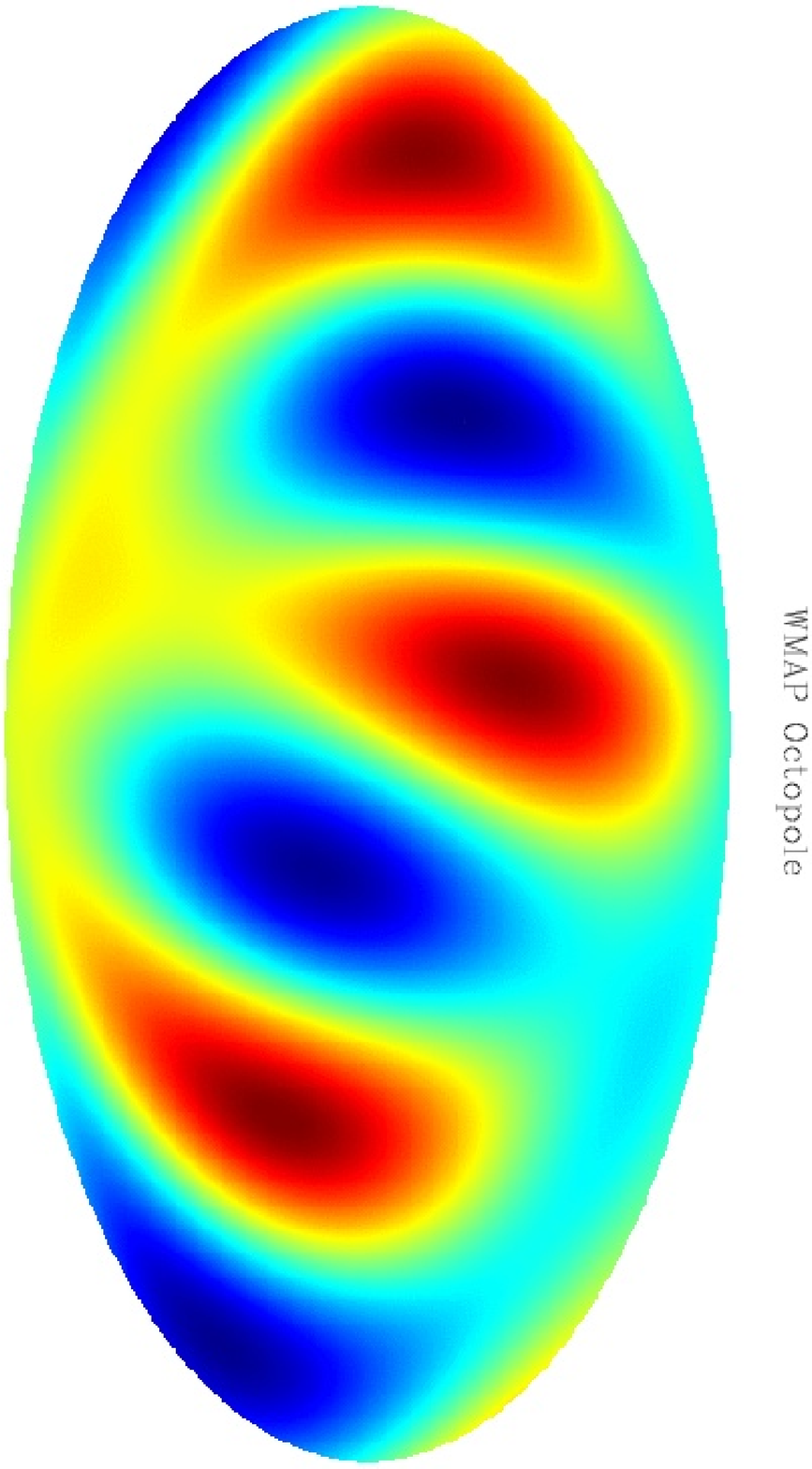}
 \includegraphics[width=5cm,angle=90]{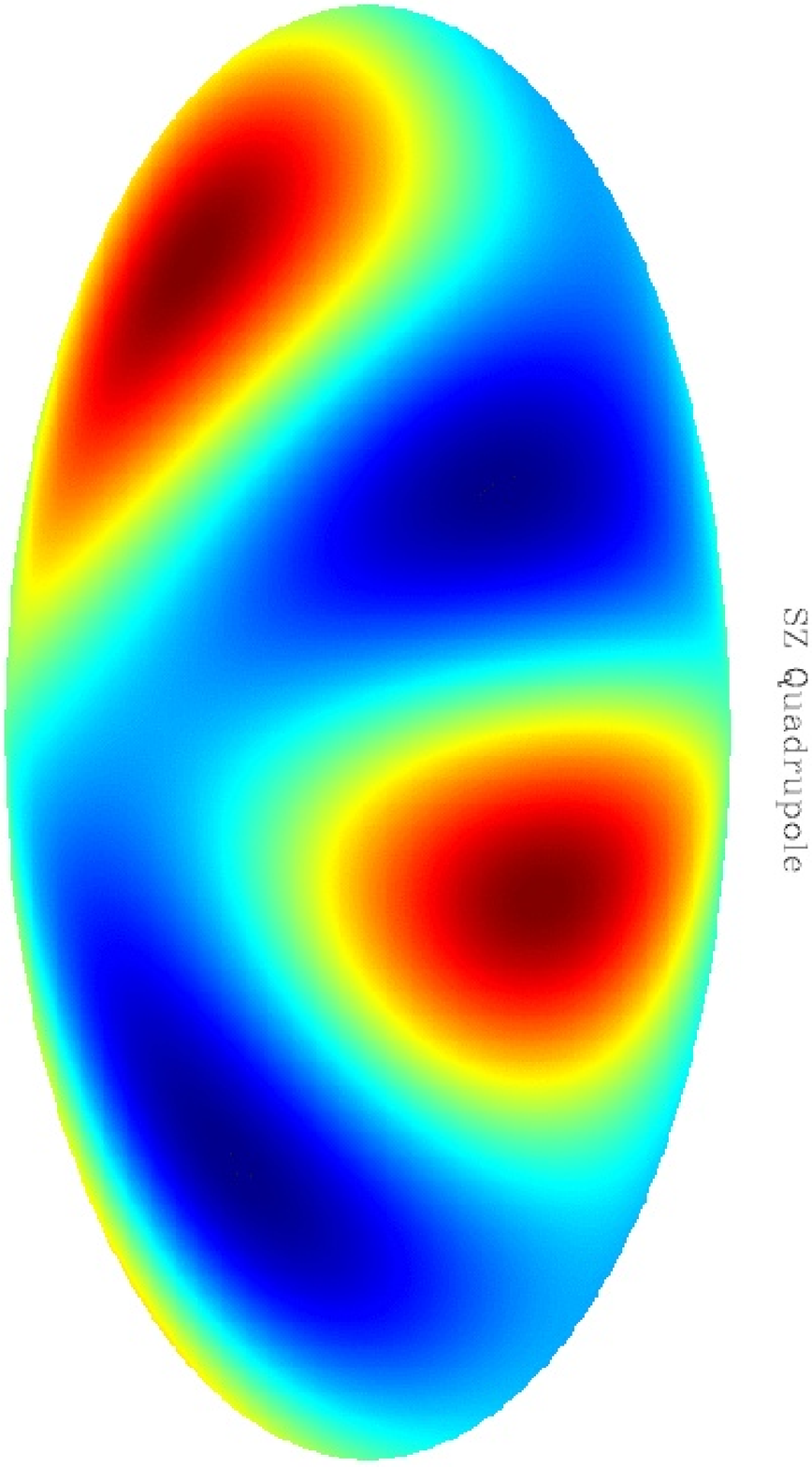}
 \includegraphics[width=5cm,angle=90]{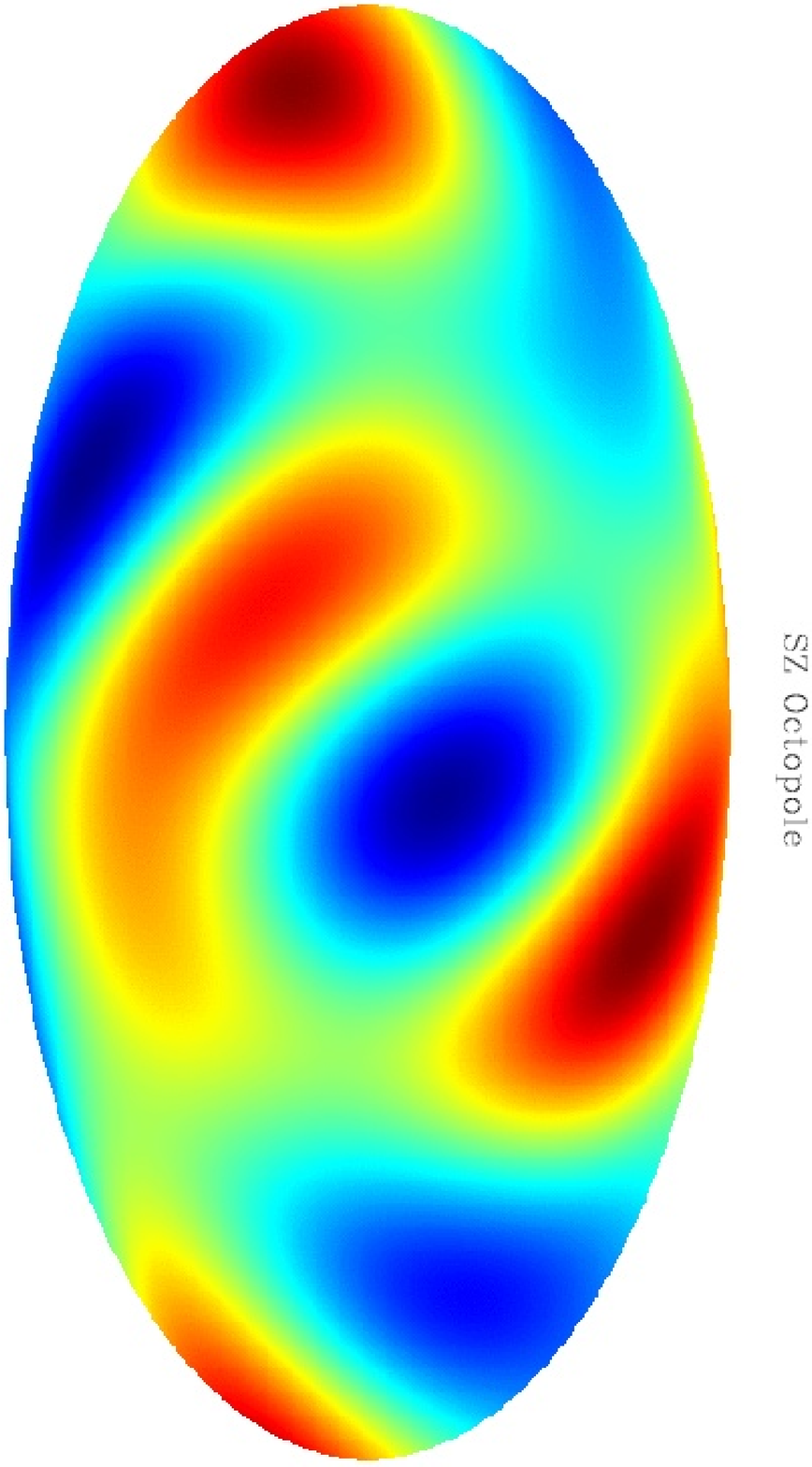}
\caption{Comparison between the quadrupole (left panels) and octopole
(right panels) obtained from the WMAP internal linear combination
(ILC) map (upper panels) and from SZ template extracted from the {\it
gas} simulation (lower panels).}
\label{fig:lowl}
\end{center}
\end{figure*}

Several anomalies have been detected at low multipoles in the WMAP
data. In particular \cite{tegmark1} and \cite{tegmark2} find the CMB
quadrupole and octopole to have an unusually high degree of alignment,
which is significant at the $2\sigma$ level.  Moreover they show that
the directions of the quadrupole and the octopole are close to the
direction of the Virgo cluster which could suggest a possible link to
structures in the local universe. By analysing our SZ templates, we
find that the directions of the quadrupole and octopole (shown in
Fig. \ref{fig:lowl} for the {\it gas} simulation) are very different
from that extracted from the WMAP data: in our maps the quadrupole and
octopole point toward $(b,l)
\approx (42.3^\circ,-177.6^\circ)$ and $(b,l) 
\approx (32.5^\circ,88.1^\circ)$, respectively, 
while the preferred axes in the WMAP data are in the directions $(b,l)
\approx (58.8^\circ,-102.4^\circ)$ for the quadrupole and $(b,l) \approx
(62.0^\circ,-121.6^\circ)$ for the octopole.  Moreover the amplitudes
of both quadrupole and octopole are far too small to have such an
effect.

Finally it has been claimed by different groups that the quadrupole of
the CMB is anomalously low [see the discussion in \cite{efst} and
references therein] and that this could be caused by structures
present in the local universe \citep{brasiliani}. Again, we find that
the amplitude of the SZ effect is far too low to have any relevant
effect.

\subsection{Detectability of the local SZ signal with the Planck satellite}

We investigate now whether the local SZ effect can be observed using
the upcoming Planck satellite. The very low noise level and the huge
frequency range of the Planck experiment could make it capable of
seeing this effect at large scales. Note that the CMB itself is the
most important contamination when trying to look for the SZ effect at
large scales \citep{hansen}. As the CMB has the same temperature at
all frequencies, using a map produced by computing the difference
between two frequency bands could be a promising way of detecting the
SZ effect \citep[this strategy was attempted for the WMAP data in
][]{hansen}.  This makes the observation of the SZ effect limited only
by the instrumental noise. Unfortunately galactic foregrounds also
have several frequency dependent components, and these need to be
eliminated at a very high degree in order not to confuse detections of
the SZ signal. Here we will assume that these have been completely
removed outside a conservative galactic cut.

In order to estimate the sensitivity to the local SZ effect for the
Planck experiment, we simulated Planck data for the 70, 100, 143 and
217 GHz channels, by using the technical specifications given at the
Planck home
page\footnote{http://www.rssd.esa.int/index.php?project=PLANCK}.  In
more detail, we apply the following procedure.
\begin{itemize}
\item 
For each simulation, we generate a CMB sky and smooth it by using the
beams corresponding to the 4 channels of interest. The smoothing
is performed in the harmonic space by using beams with size of 14, 9.5,
7.1 and 5 arcmin for the channels at 70, 100, 143 and 217 Ghz,
respectively. As the band width for each frequency channel is expected
to be much smaller than the frequency difference between the channels,
we approximate the band width to be infinitely narrow
\item 
We generate white, non-uniform noise for each channel and add it to
the maps. The Planck experiment is expected to have highly correlated
noise which would affect the noise power spectrum at the largest
scales. Since here we are only interested in a rough estimate of the
sensitivity of the Planck experiment to the SZ effect, we neglect this
effect. Consequently our results could be slightly over-optimistic at
the largest scales.
\item 
We add the SZ template to each map by using the frequency dependence
of the SZ effect.
\item 
We construct three different maps, corresponding to the difference
between the maps at 217 and 100 GHz (called map A), between 143 and
100 GHz (map B) and between 143 and 70 GHz (map C).
\item 
We apply the Kp0 galaxy and point source mask used by the WMAP team
(and publicly available at the Lambda 
website\footnote{http://lambda.gsfc.nasa.gov/}) 
before calculating the power spectrum of the difference maps.
\end{itemize}

\begin{figure}
  \begin{center} \includegraphics[width=8cm]{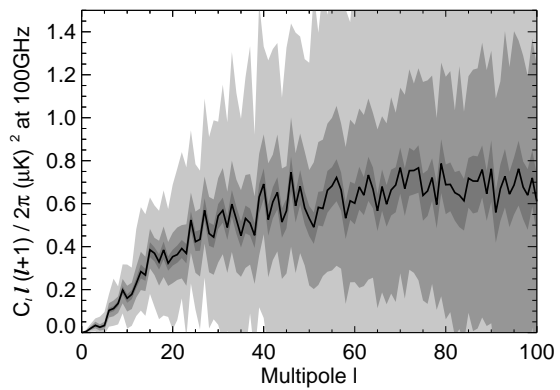}
  \end{center} \caption{ The estimated power spectra of three
  difference maps from Planck simulations. The solid line shows the
  power spectrum of the local SZ template. The shaded bands correspond
  to the $1\sigma$ spread of the estimated power spectra in 300
  simulations for the following difference maps: A = 217-100 GHz, B =
  143-100 GHz and C = 143-70 GHz. Map A has the lowest spread and map
  C has the largest one. The Kp0 galactic cut has been used in these
  simulations. Note that all spectra are normalized to the expected SZ
  effect at 100 GHz.
  \label{fig:sz_with_planck} }
\end{figure}

In Fig. \ref{fig:sz_with_planck} we show the results of a set of 300
different simulations (the number of simulations has been chosen
to obtain error bars correct within a few percent).  The solid line
corresponds to the power spectrum of the template outside the Kp0
galactic cut. The three shaded areas refer to the $1\sigma$ spread
from simulations of the three difference maps, map A having the
smallest spread, and map C the largest one. The huge difference
between the different pairs comes from two different factors: first,
the bigger the frequency difference, the bigger is the SZ effect in
the resulting difference map; second, as the CMB is absent in the
difference maps, the limiting factor is the noise level which is
highly different in different frequency channels. We notice that
assuming white noise and a perfect foreground subtraction outside the
Kp0 cut, we can expect the local SZ effect to be detected by the
Planck satellite.  Even if real error bars might be larger than our
simulations show, one will still be able to compare data with the
predicted local SZ effect. Finally note that analyzing difference maps
is a useful tool for checking that foregrounds and other systematic
effects of the experiment are well understood. For this purpose it
will be of high importance to know well the expected local SZ effect
as it will give an important contribution to the difference maps.

\section{Conclusions}

We used two hydrodynamical cosmological simulations which are designed
to represent the observed large-scale structure of the local universe
(up to approximately 110 Mpc from the Milky Way) to investigate the
imprints of the extended SZ signal caused by local superclusters onto
the cosmological full-sky signal. The two simulations, which assume
the standard $\Lambda$CDM model, started from the same initial
conditions, but included the treatment of a different set of physical
processes: the first one assumes adiabatic physics only, while the
second one follows also cooling, star formation and supernovae
feedback.

We find that the largest and most prominent structures observed in the
SZ map constructed from our constrained simulations are caused by
local superclusters like Pisces-Perseus or the Centaurus supercluster
region.  Even Virgo, a relatively poor but nearby cluster, belongs to
the most prominent and extended features of the map. Comparing the
power spectra of the thermal SZ signal with the one derived from the
{\it Hubble volume simulation} \citep{schaefer2004a}, we find that for
$\ell<100$ these structures lead to an amplitude which is at least
twice as high (one order of magnitude for $\ell<20$).  When comparing
the derived power spectra for the kinetic SZ this factor is even
higher.

However, the overall amplitude of the effect is far too small (3-4
orders of magnitude) to have any significant influence on the lowest
multipoles of the CMB.  There have been some claims that the
anomalously low CMB quadrupole and octopole observed by the WMAP
satellite could be explained by the imprints of local superclusters.
We have shown that the amplitude of the SZ low multipoles must be
considerably much higher in order to produce such an effect as well as
to have an influence on the observed quadrupole and octopole.
Moreover, the directions on the sky of the quadrupole and octopole
inferred from our simulations do not coincide with the observed
direction in the WMAP maps.

Both for our {\it gas} simulation as well as for the {\it Hubble
volume simulation}, radiative cooling processes and feedback by star
formation have been neglected, which could change the absolute
amplitude of the inferred SZ signal of galaxy clusters.  However,
using a simulation which includes all these effects (the {\it csf}
simulation), we even find a slight decrease of the SZ signal, similar
to the findings obtained by using isolated cluster simulations. This
will make it even more unlikely to explain the abnormal lower
multipoles observed in the WMAP data with the SZ signal of the local
universe.

Nevertheless, we find that when taking advantage of the huge frequency
range and low noise level of the Planck satellite, an estimate of the
SZ power spectrum at large scales could be obtained. We have shown
this by producing a set of simulated CMB maps with experimental noise
at several frequencies. Analyzing the power spectrum of maps obtained
by taking the difference of CMB maps at different frequencies, we show
that the power spectrum of the SZ effect is limited only by
experimental noise and foreground contamination. We have found that
the predicted noise level of the Planck satellite is low enough to
allow for the detection of the local SZ power spectrum at low
multipoles provided that the galaxy can be removed with sufficient
accuracy. It is out of the scope of this work to obtain an accurate
estimate of the predicted detection level by the Planck satellite, for
which correlated noise as well as an accurate study of foreground
removal procedures is required.

\section*{Acknowledgements}

We would like to warmly thank Luca Tornatore for providing the
self-consistent feedback scheme handling the metal production and
metal cooling, and for his assistance when performing the {\it csf}
run.  We are also grateful to Bjoern Sch\"afer for providing the SZ
maps derived from the {\it Hubble volume simulation} and the MPAC team
at MPA for the access to analysis software.  We would like also thank
A. Banday for a critical reading of the manuscript and useful
discussions.  The hydrodynamical simulations presented in this work
have been performed using computer facilities at the Rechenzentrum der
Max-Planck-Gesellschaft, Garching (the {\it gas} simulation), and at
the University of Tokyo supported by the Special Coordination Fund for
Promoting Science and Technology, Ministry of Education, Culture,
Sport, Science and Technology (the {\it csf} simulation).  KD
acknowledges partial support by a Marie Curie Fellowship of the
European Community program ``Human Potential'' under contract number
MCFI-2001-01227. FKH acknowledges financial support from the Norwegian
Research Council. We acknowledge use of the
\hp~ \citep{healpix} software and analysis package for deriving the
results in this paper.

\bibliographystyle{mn2e}

\bibliography{master}

\begin{thebibliography}{}

\bibitem[\protect\citeauthoryear{Abramo \& Sodre}{Abramo \&
  Sodre}{2003}]{brasiliani}
Abramo R.,  Sodre J.,  2003, preprint, astro-ph/0312124

\bibitem[\protect\citeauthoryear{{Afshordi}, {Loh} \& {Strauss}}{{Afshordi}
  et~al.}{2004}]{afshordi2004b}
{Afshordi} N.,  {Loh} Y.,    {Strauss} M.~A.,  2004, Phys. Rev. D, 69, 083524

\bibitem[\protect\citeauthoryear{{Bartelmann}}{{Bartelmann}}{2001}]{bartelmann%
2001}
{Bartelmann} M.,  2001, \aap, 370, 754

\bibitem[\protect\citeauthoryear{{Battistelli}, {De Petris}, {Lamagna}, {Luzzi}
  et~al.,}{{Battistelli} et~al.}{2003}]{2003ApJ...598L..75B}
{Battistelli} E.~S.,  {De Petris} M.,  {Lamagna} L.,  {Luzzi} G.,    et~al.,
  2003, \apj, 598, L75

\bibitem[\protect\citeauthoryear{Bennett, Bay, Halpern, Hinshaw
  et~al.,}{Bennett et~al.}{2003}]{BE03.1}
Bennett C.,  Bay M.,  Halpern M.,  Hinshaw G.,    et~al., 2003, ApJ, 583, 1

\bibitem[\protect\citeauthoryear{{Colberg}, {White}, {Yoshida}, {MacFarland}
  et~al.,}{{Colberg} et~al.}{2000}]{2000MNRAS.319..209C}
{Colberg} J.~M.,  {White} S.~D.~M.,  {Yoshida} N.,  {MacFarland} T.~J.,
  et~al., 2000, \mnras, 319, 209

\bibitem[\protect\citeauthoryear{{Cooray}, {Baumann} \& {Sigurdson}}{{Cooray}
  et~al.}{2004}]{cooray2004}
{Cooray} A.,  {Baumann} D.,    {Sigurdson} K.,  2004, preprint,
  astro-ph/0410006

\bibitem[\protect\citeauthoryear{{da Silva}, {Kay}, {Liddle} \& {Thomas}}{{da
  Silva} et~al.}{2004}]{dasilva2004}
{da Silva} A.~C.,  {Kay} S.~T.,  {Liddle} A.~R.,    {Thomas} P.~A.,  2004,
  \mnras, 348, 1401

\bibitem[\protect\citeauthoryear{de Oliveira-Costa, Tegmark, Zaldarriaga \&
  Hamilton}{de~Oliveira-Costa et~al.}{2004}]{tegmark2}
de Oliveira-Costa A.,  Tegmark M.,  Zaldarriaga M.,    Hamilton A.,  2004,
  Phys. Rev. D, 69, 063516

\bibitem[\protect\citeauthoryear{{Diego}, {Silk} \& {Sliwa}}{{Diego}
  et~al.}{2003}]{diego2003}
{Diego} J.~M.,  {Silk} J.,    {Sliwa} W.,  2003, New Astronomy Review, 47, 855

\bibitem[\protect\citeauthoryear{{Dolag}, {Grasso}, {Springel} \&
  {Tkachev}}{{Dolag} et~al.}{2005}]{2005JCAP...01..009D}
{Dolag} K.,  {Grasso} D.,  {Springel} V.,    {Tkachev} I.,  2005, JCAP, 1, 9

\bibitem[\protect\citeauthoryear{Efstathiou}{Efstathiou}{2004}]{efst}
Efstathiou G.,  2004, MNRAS, 348, 885

\bibitem[\protect\citeauthoryear{{Efstathiou}}{{Efstathiou}}{2004}]{efsta2004}
{Efstathiou} G.,  2004, \mnras, 348, 885

\bibitem[\protect\citeauthoryear{{Fixsen}}{{Fixsen}}{2003}]{2003ApJ...594L..67%
F}
{Fixsen} D.~J.,  2003, \apjl, 594, L67

\bibitem[\protect\citeauthoryear{{Fosalba} \& {Gazta{\~ n}aga}}{{Fosalba} \&
  {Gazta{\~ n}aga}}{2004}]{fosalba2004}
{Fosalba} P.,  {Gazta{\~ n}aga} E.,  2004, \mnras, 350, L37

\bibitem[\protect\citeauthoryear{{Gazta{\~ n}aga}, {Wagg}, {Multam{\" a}ki},
  {Monta{\~ n}a} \& {Hughes}}{{Gazta{\~ n}aga} et~al.}{2003}]{gazta2003}
{Gazta{\~ n}aga} E.,  {Wagg} J.,  {Multam{\" a}ki} T.,  {Monta{\~ n}a} A.,
  {Hughes} D.~H.,  2003, \mnras, 346, 47

\bibitem[\protect\citeauthoryear{G\'orski, Hivon \& Wandelt}{G\'orski
  et~al.}{1998}]{healpix}
G\'orski K.~M.,  Hivon E.,    Wandelt B.~D.,  1998, `Analysis Issues for Large
  CMB Data Sets', 1998, eds A. J. Banday,R. K. Sheth and L. Da Costa, ESO,
  Printpartners Ipskamp, NL, pp.37-42 (astro-ph/9812350); Healpix HOMEPAGE:
  http://www.eso.org/science/healpix/

\bibitem[\protect\citeauthoryear{Hansen, Branchini, Mazzotta, Cabella \&
  Dolag}{Hansen et~al.}{2005}]{hansen}
Hansen F.~K.,  Branchini E.,  Mazzotta P.,  Cabella P.,    Dolag K.,  2005,
  MNRAS, submitted, astro-ph/0502227

\bibitem[\protect\citeauthoryear{{Hern{\' a}ndez-Monteagudo}, {Genova-Santos}
  \& {Atrio-Barandela}}{{Hern{\' a}ndez-Monteagudo}
  et~al.}{2004}]{2004ApJ...613L..89H}
{Hern{\' a}ndez-Monteagudo} C.,  {Genova-Santos} R.,    {Atrio-Barandela} F.,
  2004, \apjl, 613, L89

\bibitem[\protect\citeauthoryear{{Hernandez-Monteagudo}, {Genova-Santos} \&
  {Atrio-Barandela}}{{Hernandez-Monteagudo} et~al.}{2004}]{hern2004b}
{Hernandez-Monteagudo} C.,  {Genova-Santos} R.,    {Atrio-Barandela} F.,  2004,
  preprint, astro-ph/0406428

\bibitem[\protect\citeauthoryear{{Hirata}, {Padmanabhan}, {Seljak}, {Schlegel}
  \& {Brinkmann}}{{Hirata} et~al.}{2004}]{hirata2004}
{Hirata} C.~M.,  {Padmanabhan} N.,  {Seljak} U.,  {Schlegel} D.,    {Brinkmann}
  J.,  2004, Phys. Rev. D, 70, 103501

\bibitem[\protect\citeauthoryear{{Hoffman} \& {Ribak}}{{Hoffman} \&
  {Ribak}}{1991}]{Hoffman1991}
{Hoffman} Y.,  {Ribak} E.,  1991, \apj, 380, L5

\bibitem[\protect\citeauthoryear{{Huffenberger}, {Seljak} \&
  {Makarov}}{{Huffenberger} et~al.}{2004}]{huffenberger2004}
{Huffenberger} K.~M.,  {Seljak} U.,    {Makarov} A.,  2004, Phys. Rev. D, 70,
  063002

\bibitem[\protect\citeauthoryear{{Ikebe}, {Reiprich}, {B{\" o}hringer},
  {Tanaka} \& {Kitayama}}{{Ikebe} et~al.}{2002}]{2002A&A...383..773I}
{Ikebe} Y.,  {Reiprich} T.~H.,  {B{\" o}hringer} H.,  {Tanaka} Y.,
  {Kitayama} T.,  2002, \aap, 383, 773

\bibitem[\protect\citeauthoryear{{Jenkins}, {Frenk}, {White}, {Colberg},
  {Cole}, {Evrard}, {Couchman} \& {Yoshida}}{{Jenkins}
  et~al.}{2001}]{2001MNRAS.321..372J}
{Jenkins} A.,  {Frenk} C.~S.,  {White} S.~D.~M.,  {Colberg} J.~M.,  {Cole} S.,
  {Evrard} A.~E.,  {Couchman} H.~M.~P.,    {Yoshida} N.,  2001, \mnras, 321,
  372

\bibitem[\protect\citeauthoryear{{Kolatt}, {Dekel}, {Ganon} \&
  {Willick}}{{Kolatt} et~al.}{1996}]{Kolatt:1996}
{Kolatt} T.,  {Dekel} A.,  {Ganon} G.,    {Willick} J.~A.,  1996, \apj, 458,
  419

\bibitem[\protect\citeauthoryear{{Mathis}, {Lemson}, {Springel}, {Kauffmann},
  {White}, {Eldar} \& {Dekel}}{{Mathis} et~al.}{2002}]{Mathis:2002}
{Mathis} H.,  {Lemson} G.,  {Springel} V.,  {Kauffmann} G.,  {White} S.~D.~M.,
  {Eldar} A.,    {Dekel} A.,  2002, \mnras, 333, 739

\bibitem[\protect\citeauthoryear{{Mazzotta}, {Rasia}, {Moscardini} \&
  {Tormen}}{{Mazzotta} et~al.}{2004}]{mazzotta:2004}
{Mazzotta} P.,  {Rasia} E.,  {Moscardini} L.,    {Tormen} G.,  2004, \mnras,
  354, 10

\bibitem[\protect\citeauthoryear{{Mohr}, {Mathiesen} \& {Evrard}}{{Mohr}
  et~al.}{1999}]{1999ApJ...517..627M}
{Mohr} J.~J.,  {Mathiesen} B.,    {Evrard} A.~E.,  1999, \apj, 517, 627

\bibitem[\protect\citeauthoryear{{Monaghan} \& {Lattanzio}}{{Monaghan} \&
  {Lattanzio}}{1985}]{1985A&A...149..135M}
{Monaghan} J.~J.,  {Lattanzio} J.~C.,  1985, \aap, 149, 135

\bibitem[\protect\citeauthoryear{{Motl}, {Hallman}, {Burns} \& {Norman}}{{Motl}
  et~al.}{2005}]{motl2005}
{Motl} P.~M.,  {Hallman} E.~J.,  {Burns} J.~O.,    {Norman} M.~L.,  2005, \apj,
  623, L63

\bibitem[\protect\citeauthoryear{{Myers}, {Shanks}, {Outram}, {Frith} \&
  {Wolfendale}}{{Myers} et~al.}{2004}]{myers2004}
{Myers} A.~D.,  {Shanks} T.,  {Outram} P.~J.,  {Frith} W.~J.,    {Wolfendale}
  A.~W.,  2004, \mnras, 347, L67

\bibitem[\protect\citeauthoryear{{Rasia}, {Mazzotta}, {Borgani}, {Moscardini},
  {Dolag}, {Tormen}, {Diaferio} \& {Murante}}{{Rasia}
  et~al.}{2005}]{rasia:2005}
{Rasia} E.,  {Mazzotta} P.,  {Borgani} S.,  {Moscardini} L.,  {Dolag} K.,
  {Tormen} G.,  {Diaferio} A.,    {Murante} G.,  2005, \apj, 618, L1

\bibitem[\protect\citeauthoryear{{Salpeter}}{{Salpeter}}{1955}]{1955ApJ...121.%
.161S}
{Salpeter} E.~E.,  1955, \apj, 121, 161

\bibitem[\protect\citeauthoryear{{Sanderson}, {Ponman}, {Finoguenov},
  {Lloyd-Davies} \& {Markevitch}}{{Sanderson}
  et~al.}{2003}]{2003MNRAS.340..989S}
{Sanderson} A.~J.~R.,  {Ponman} T.~J.,  {Finoguenov} A.,  {Lloyd-Davies} E.~J.,
     {Markevitch} M.,  2003, \mnras, 340, 989

\bibitem[\protect\citeauthoryear{{Schaefer}, {Pfrommer}, {Bartelmann},
  {Springel} \& {Hernquist}}{{Schaefer} et~al.}{2004}]{schaefer2004a}
{Schaefer} B.~M.,  {Pfrommer} C.,  {Bartelmann} M.,  {Springel} V.,
  {Hernquist} L.,  2004, preprint, astro-ph/0407089

\bibitem[\protect\citeauthoryear{{Spergel}, {Verde}, {Peiris}, {Komatsu}
  et~al.,}{{Spergel} et~al.}{2003}]{SP03.1}
{Spergel} D.~N.,  {Verde} L.,  {Peiris} H.~V.,  {Komatsu} E.,    et~al., 2003,
  \apjs, 148, 175

\bibitem[\protect\citeauthoryear{{Springel}}{{Springel}}{2005}]{springel2005}
{Springel} V.,  2005, preprint, astro-ph/0505010

\bibitem[\protect\citeauthoryear{{Springel} \& {Hernquist}}{{Springel} \&
  {Hernquist}}{2002}]{2002MNRAS.333..649S}
{Springel} V.,  {Hernquist} L.,  2002, \mnras, 333, 649

\bibitem[\protect\citeauthoryear{{Springel} \& {Hernquist}}{{Springel} \&
  {Hernquist}}{2003}]{2003MNRAS.339..289S}
{Springel} V.,  {Hernquist} L.,  2003, \mnras, 339, 289

\bibitem[\protect\citeauthoryear{Springel, Yoshida \& White}{Springel
  et~al.}{2001}]{SP01.1}
Springel V.,  Yoshida N.,    White S.,  2001, New Astronomy, 6, 79

\bibitem[\protect\citeauthoryear{{Sunyaev} \& {Zeldovich}}{{Sunyaev} \&
  {Zeldovich}}{1972}]{sz}
{Sunyaev} R.~A.,  {Zeldovich} Y.~B.,  1972, Comments on Astrophysics and Space
  Physics, 4, 173

\bibitem[\protect\citeauthoryear{Tegmark, de Oliveira-Costa \&
  Hamilton}{Tegmark et~al.}{2003}]{tegmark1}
Tegmark M.,  de Oliveira-Costa A.,    Hamilton A.,  2003, Phys. Rev. D, 68,
  123523

\bibitem[\protect\citeauthoryear{{Tornatore}, {Borgani}, {Matteucci}, {Recchi}
  \& {Tozzi}}{{Tornatore} et~al.}{2004}]{2004MNRAS.349L..19T}
{Tornatore} L.,  {Borgani} S.,  {Matteucci} F.,  {Recchi} S.,    {Tozzi} P.,
  2004, \mnras, 349, L19

\bibitem[\protect\citeauthoryear{{Tornatore}, {Borgani}, {Springel},
  {Matteucci}, {Menci} \& {Murante}}{{Tornatore} et~al.}{2003}]{tornatore2003}
{Tornatore} L.,  {Borgani} S.,  {Springel} V.,  {Matteucci} F.,  {Menci} N.,
  {Murante} G.,  2003, \mnras, 342, 1025

\bibitem[\protect\citeauthoryear{{Vikhlinin}}{{Vikhlinin}}{2005}]{vikhlinin:20%
05}
{Vikhlinin} A.,  2005, preprint, astro-ph/0504098

\bibitem[\protect\citeauthoryear{{White}, {Hernquist} \& {Springel}}{{White}
  et~al.}{2002}]{2002ApJ...579...16W}
{White} M.,  {Hernquist} L.,    {Springel} V.,  2002, \apj, 579, 16

\end{thebibliography}
\label{lastpage}
\end{document}